\documentclass[fleqn,usenatbib]{mnras}

\usepackage{newtxtext,newtxmath}

\usepackage[T1]{fontenc}
\usepackage{ae,aecompl}
\usepackage{multirow}
\usepackage{longtable}

\usepackage[pdftex]{graphicx}	
\usepackage{amsmath}	
\usepackage{CJKutf8}
\usepackage{makecell}

\newcommand{\dd}{{\rm d}}
\newcommand{\REarth}{$R_\oplus$}
\newcommand{\MEarth}{$M_\oplus$}
\newcommand{\water}{H$_{2}$O}

\title[Water Equation of State]{Implications of an improved water equation of state for water-rich planets}

\author[Huang et al.]{Chenliang~Huang(黄辰亮),$^{1,2}$ David~R.~Rice,$^{1}$ Zachary~M.~Grande,$^{3}$ Dean~Smith$^{3,5}$, \newauthor Jesse~S.~Smith,$^{5}$ John~H~Boisvert,$^{1}$ Oliver~Tschauner,$^{4}$ Ashkan~Salamat,$^{3}$  \newauthor and Jason~H.~Steffen$^{1}$\\
\\
\normalsize{$^{1}$Department of Physics and Astronomy, University of Nevada Las Vegas, Las Vegas, Nevada 89154, USA}\\
\normalsize{$^{2}$Lunar and Planetary Laboratory, University of Arizona, Tucson, Arizona 85721, USA}\\
\normalsize{$^{3}$Department of Physics and Astronomy and HiPSEC, University of Nevada Las Vegas, Las Vegas, Nevada 89154, USA}\\
\normalsize{$^{4}$Department of Geoscience, University of Nevada Las Vegas, Las Vegas, Nevada 89154, USA}\\
\normalsize{$^{5}$HPCAT, X-ray Science Division, Argonne National Laboratory, Argonne, Illinois 60439, USA}
}

\date{Accepted XXX. Received YYY; in original form ZZZ}

\pubyear{2019}

\begin{document}
\begin{CJK*}{UTF8}{gbsn}
\label{firstpage}
\pagerange{\pageref{firstpage}--\pageref{lastpage}}
\maketitle
\end{CJK*}

\begin{abstract}
Water (H$_{2}$O), in all forms, is an important constituent in planetary bodies, controlling habitability and influencing geological activity.  Under conditions found in the interior of many planets, as the pressure increases, the H-bonds in water gradually weaken and are replaced by ionic bonds.  Recent experimental measurements of the water equation of state (EOS) showed both a new phase of H-bonded water ice, ice-VII$_{\text{t}}$, and a relatively low transition pressure just above 30\,GPa to ionic bonded ice-X, which has a bulk modulus 2.5 times larger.  The higher bulk modulus of ice-X produces larger planets for a given mass, thereby either reducing the atmospheric contribution to the volume of many exoplanets or limiting their water content.  We investigate the impact of the new EOS measurements on the planetary mass-radius relation and interior structure for water-rich planets.  We find that the change in the planet mass-radius relation caused by the systematic differences between previous and new experimental EOS measurements are comparable to the observational uncertainties in some planet sizes---an issue that will become more important as observations continue to improve.  
\end{abstract}

\begin{keywords}
planets and satellites: interiors, composition -- methods: numerical
\end{keywords}



\section{Introduction}\label{sec:intro}

Water, H$_2$O, is a fundamental building block of planets -- being composed of two of the three most abundant elements in the universe.  Water also has complex solid-state properties as the nature of the intermolecular bonds change with pressure and temperature.  Measurements of the properties of water under conditions relevant to planetary interiors suffer from several systematic effects that have limited our understanding of its crystalline structure, compressibility, equation of state (EOS), and phase boundaries.  For example, the phase boundary between ice-VII and ice-X (the two dominant structures of ice above room temperature) from various studies are inconsistent and give a broad range for the transition pressure that spans from 40 GPa to beyond 120 GPa
\citep{hirsch1986effect,guthrie2019structure}.

Historically, the EOS of water-ice systems has been determined by compression of the sample with limited ability to address issues of non-hydrostaticity and the deviatoric stress that arises from high pressure methods associated with diamond anvil cell techniques \citep[e.g.][]{Frank2004}.  Recent measurements of the water EOS by \citet{Grande:2019} applied new experimental techniques to alleviate some of these systematic effects.  
They used a mid-IR laser system to directly heat a water sample within a diamond anvil cell, melting and then recrystallizing the sample prior to making x-ray diffraction measurements of the structure.  This ``normalization'' process provides two major benefits.  First, it produces many small crystalline domains rather than a few large ones.  These smaller domains yield higher quality Debye--Sherrer fringes that are more amenable to standard analysis methods.  Second, it alleviates deviatoric stresses that distort the x-ray signature in cell volume measurements. 

Two key results from the \citet{Grande:2019} work include: 1) the discovery, near 5 GPa, of a new phase of ice, denoted ice-VII$_{\text{t}}$, which has a tetragonal crystalline structure rather than the cubic structure of neighbouring phases, and 2) an improved measurement of the pressure (near 31 GPa) where covalently bonded ice-VII transforms to the ionic bonded ice-X.  Their measurements have important implications for the properties of water-rich planets -- including bulk properties such as the mass-radius relationship and the fundamental nature of water trapped below the surface.  In this work, we investigate how this new EOS could change the inferred bulk composition and structure of observed exoplanets.

The planet interior model, which relies on the EOSs of its various components, can be used to infer from its observed properties various aspects of planet composition such as water content \citep[e.g.][]{Unterborn2018NA, Unterborn2018RNAAS} and atmosphere mass \citep{TOI-849}.  However, \citet{Hakim18} showed that the modelling uncertainty that arises from different EOSs of iron are larger than the observational uncertainties for the best-observed super-Earths.  Thus, improved EOS measurements of key planet-building materials are essential for understanding the nature of observed exoplanets.  For H$_2$O specifically, many experimental and theoretical studies of the EOS across phase space (e.g. \citet{French2009,Mazevet2019,Journaux2020}) were recently compiled in \citet{Haldemann2020}.

Below, we introduce a planet interior model using the new high-pressure ice EOS measurement from \citet{Grande:2019}.  We show the interior structure of water-rich planets based upon our model.  And, we indicate how these measurements affect the mass-radius relationship and inferences of planet composition.  We conclude by discussing other ramifications of these results for planet properties and encouraging future studies of the water EOS.

\section{Planet model and Equations of State of Ice}
\label{sec:model}

For our planet structure modelling, we consider a fully differentiated, spherically symmetric planet composed of two distinct layers\footnote{This is a simplified solver and we present our full planet modeling code which includes up to four layers in \citet{Magrathea}.  This code will be located at https://github.com/Huang-CL/Magrathea.}.   We calculate the mass $m(r)$ within radius $r$, pressure $P(r)$ and density $\rho(r)$ at radius $r$ by solving the following three equations.

\begin{enumerate}
\item Mass continuity equation
\begin{equation}
  \label{eq:1}
  \frac{\dd m(r)}{\dd r} = 4 \pi r^2 \rho(r),
\end{equation}

\item hydrostatic equilibrium
\begin{equation}
  \label{eq:2}
  \frac{\dd P(r)}{\dd r} = \frac{-G m(r) \rho(r)}{r^2},
\end{equation}

\item equation of state
\begin{equation}
  \label{eq:3}
  \rho(r) = f(P).
\end{equation}
\end{enumerate}

\noindent
Our code integrates ordinary differential equations inside-out from an initial guess for the central pressure.  Using the shooting method, we iterate the centre pressure to match the boundary condition that the pressure at the planet surface is 1\,bar.  Given the masses, EOSs, and phase diagrams of the two components, this code yields estimates of the radius and internal structure of the planet.

\begin{figure}
\includegraphics[width=\columnwidth]{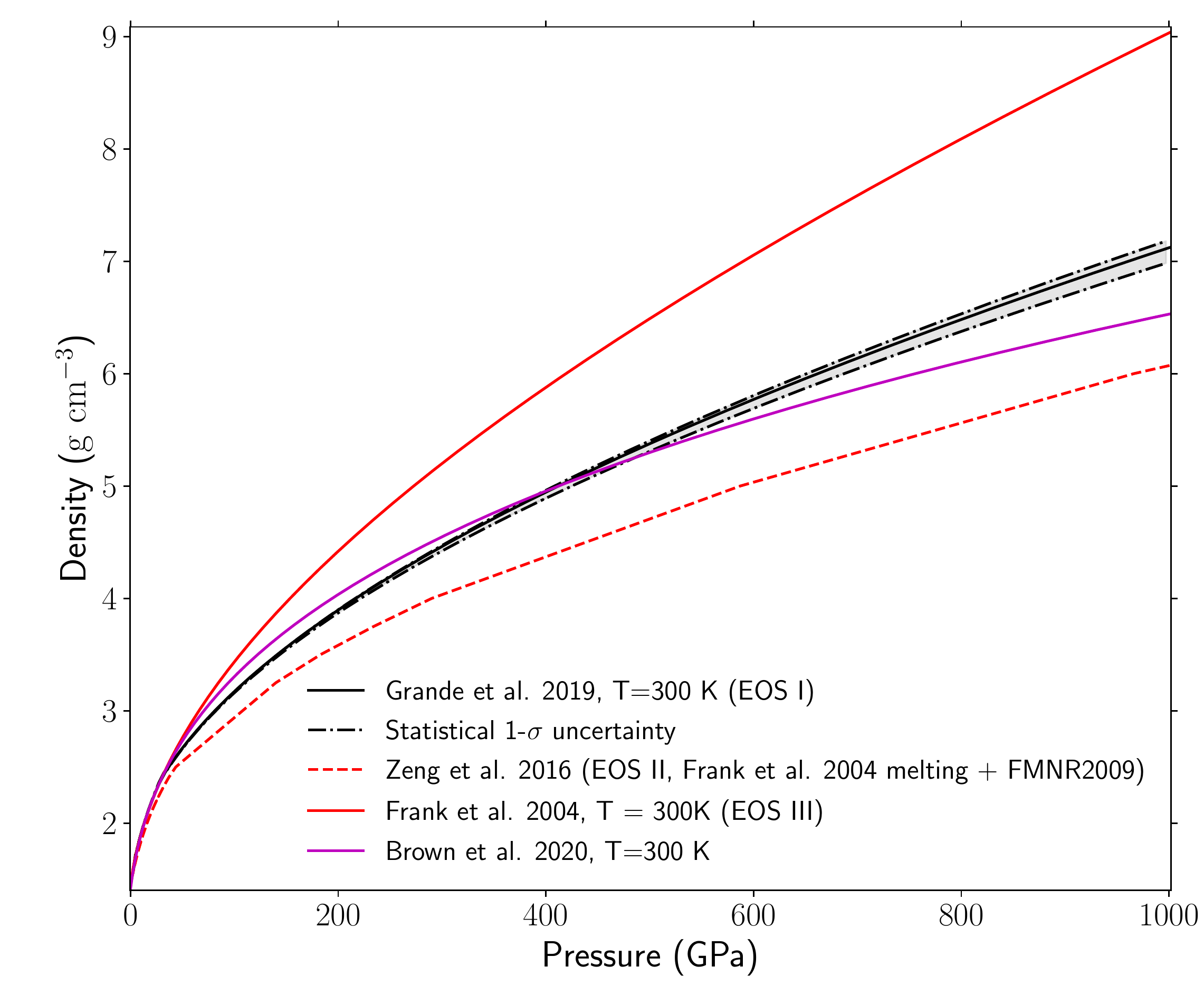}
\caption{
Comparison of the EOSs of high pressure ice.  The black solid line shows the three-phase EOS and its extrapolation suggested in~\citet{Grande:2019}.  The 1-$\sigma$ statistical uncertainty of the EOS is shown as the grey band enveloped by dash-dotted lines.  The red dashed line shows the ice EOSs used in \citet{Zeng2016} -- EOS measured by \citet{Frank2004} along the ice melting temperature for $P \leq 37.4$\,GPa and interpolated densities at a series of discrete pressure and temperature points for $P>37.4$\,GPa simulated by \citet{French2009}.  The Vinet EOS extrapolation using the $V_0$, $K_0$, and $K_0'$ fitting by \citet{Frank2004} at 300\,K is shown in the red solid curve. The EOS by \citet{brown2020local} is also shown, magenta solid curve, for comparison.
}
\label{EOS_Ice}
\end{figure}

For the ice shell of our planets, we adopt the three-phase experimental EOS in \citet{Grande:2019} (EOS~I) above the phase transition pressure from ice-VI to ice-VII at 2.1\,GPa~\citep{Dunaeva2010}.  This EOS is shown as the black solid line in Fig.~\ref{EOS_Ice}.  In order to extrapolate the ice-X EOS into the pressure regime above the maximum pressure measurement at 88\,GPa, we use the Vinet EOS since it is better behaved than the third-order Birch-Murnaghan EOS in the high pressure regime~\citep{Cohen2000}.

For our models we adopt an isothermal ice layer at 300\,K because \citet{Grande:2019} only gives the ice EOS and phase diagram at 300\,K, and because the thermal properties of high-pressure ice that are required to model the planet internal temperature profile (such as its  Gr\"{u}neisen parameter and adiabatic bulk modulus) are not well understood.  By comparison, \citet{Zeng2016} used the derived EOS for 2.22\,$\leq{}P\leq$\,37.4\,GPa from~\citet{Frank2004} along the melting line (whose temperature increases from 360~K to 866~K in this pressure range), along with an interpolated EOS from quantum molecular dynamics simulations for pressures in the range of 37.4\,GPa\,$\leq{}P\leq{}$\,8.89\,TPa from~\citet{French2009}.  Those molecular dynamics simulations calculate the EOS at a series of discrete temperature points with 1000 K intervals.  The ice density of along the 1000\,K, 2000\,K, and 4000\,K isotherms are used in the planet model.  This model is shown as the red dashed line (EOS~II) in Fig.~\ref{EOS_Ice}.  

In order to show the effect of the new ice measurements on planet structure, and to avoid the impact of the uncertain temperature profile, we choose another ice EOS measured at 300\,K for comparison.  Besides measuring the EOS along the melting line that is adopted in~\citet{Zeng2016}, \citet{Frank2004} also isothermally compressed ice at 300\,K and fitted their measured density between 6.57\,GPa  and 60.52\,GPa  with a single-phase EOS.  The Vinet EOS extrapolation, using their fitted parameters $V_0$, $K_0$, and $K_0'$, is shown as the red solid curve in Fig.~\ref{EOS_Ice} (EOS~III).  The density of ice-X given by~\citet{Grande:2019} at 60\,GPa is $\sim 5$\% smaller than the density at the same pressure and temperature from \citet{Frank2004}. When extrapolating both EOSs to the higher pressures that exist deep in a water-rich super-Earth, these density differences increase.  

In addition, we also show another theoretical 300\,K EOS presented by \citet{brown2020local} in Fig.~\ref{EOS_Ice} for comparison.  To be consistent, we extrapolate their result to high-pressure using fourth-order Vinet formula  \citep[generalized Rydberg formula in][]{Stacey_2005} that includes the second derivative of the bulk modulus $K_0''$. The resulting $\rho$-P curve broadly agrees with \citet{Grande:2019} with a maximum discrepancy of 9\% at 1000~GPa.  Although the $K_0$ obtained by \citet{brown2020local} is less than half of that obtained by \citet{Grande:2019} for ice-X, their EOS is stiffer at high-pressure because of the $K_0''$ term. Since both works suggest similar $\rho$-P relation, here we focus on the experimental result provided by \citet{Grande:2019}.

Near the surface of the ice shell, where the pressure $P\,<\,2.1$\,GPa, the exact water EOS has less impact on the planet radius because of the small thickness of the region where this phase is stable.  In this pressure range, \citet{Zeng2016} assumed \water\ was in solid phases (ice-Ih, ice-III, ice-V, and ice-VI) along its melting curve~\citep{Chaplin, Choukroun2007}.  To ensure there is no temperature inversion in our model, we apply an isothermal surface layer consisting of liquid water and ice-VI with a temperature fixed at 300\,K for all EOSs (instead of the melting temperature).

In this work, we compare the room temperature experimental high-pressure ice EOS results from \citet{Grande:2019}, to the spliced ice EOS from \citet{Zeng2016}, and the room temperature ice EOS measured by \citet{Frank2004}.  The fitting parameters for the room temperature EOSs are listed in Table~\ref{tab:formula}.  
We clarify that \citet{Frank2004} did not identify the transitions to ice-VII$_\text{t}$ or ice-X.  Thus, models that use their results are extrapolations to high pressure from the single-phase, ice-VII EOS.
\citet{Grande:2019}'s parameters show larger uncertainties than those of \citet{Frank2004}.  This arises primarily from the fact that the uncertainties in \citet{Grande:2019} were determined using the posteriors from an MCMC analysis -- including a treatment for systematic uncertainties -- while \citet{Frank2004} likely uses more standard least-squares fitting (though it isn't specified).  In addition, since the \citet{Frank2004} paper fits a single EOS, the parameters for ice-VII are determined over a broader range in pressure than the equivalent parameters from \citet{Grande:2019}.  The $1-\sigma$ uncertainty band of EOS~I (Fig.~\ref{EOS_Ice}) remains narrow even when extrapolated to high-pressure because of correlations among the uncertainties of various parameters.

\begin{table}
  \centering
  \caption{\citet{Frank2004} and \citet{Grande:2019} best fit EOS parameters.  The \citet{Zeng2016} paper uses a different parameterization for their EOS \citep[see][]{Zeng2013}.}
  \label{tab:formula}
  \begin{tabular}{ccccc}
  \hline
    Phase & Transition & $V_0$ & $K_0$ & $K_0'$ \\
     & pressure (GPa) & (cm$^3$/mol) & (GPa) & \\
    \hline
    \hline
    Frank & N/A & 12.4(0.1) & 21.1(1.3) & 4.4(0.1) \\
    ice-VII & N/A & 12.80(0.26) & 18.47(4.0) & 2.51(1.51) \\
    ice-VII$\rm_t$ & 5.10(0.5) & 12.38(0.16) & 20.76(2.46) & 4.49(0.35) \\
    ice-X & 30.9(2.9) & 10.18(0.13) & 50.52(4.16) & 4.5(0.15) \\
    \hline
  \end{tabular}
\end{table}

\section{Implications for Planet Properties}

To reveal the impact of the new EOS measurement on planet radius estimates, we reproduce the broadly adopted planet model in \citet{Zeng2016}.  Following the setup in \citet{Zeng2016}, and only varying the EOS of high-pressure ice, we calculate the mass-radius relation of a pure \water\, and a planet that is 50\%~\water\ and 50\% Rock (a water-rich world without a core).  For the rocky core of the 50\%~\water\ - 50\% Rock, we use the same extrapolated Preliminary Reference Earth Model (PREM) EOS, which models Earth's upper and lower mantle well. We do not discuss the variation of mantle EOS in this work and mixtures of rock and water \citep{Shah2020,Vazan2020} since we are focusing on the EOS of ice. \citet{Unterborn2019} shows that including more detailed upper mantle EOSs has little effect on bulk properties.

Using the MCMC posterior of the three-phase EOS fitting parameters in \citet{Grande:2019}, the 1-$\sigma$ statistical uncertainty of the fitted and extrapolated EOS is shown as the grey band enveloped by dash-dotted lines in Fig.~\ref{EOS_Ice}.  Correlation between the error bars of the EOS parameters yields a density-pressure curve with smaller uncertainty than if they were uncorrelated.  We note that, in the planet models we consider, the pressure of the ice can reach up to $\sim{}$700\,GPa at the centre of a 10 Earth-mass pure water planet -- well beyond the laboratory measurements.

According to the water phase diagram~\citep{Dunaeva2010}, water transforms from the liquid phase to ice-VI at the pressure of 0.99\,GPa  under the 300\,K temperature.  The liquid water EOS from \citet{Valencia2007} and ice-VI EOS from \citet{Bezacier2014} are applied to the corresponding regions respectively.  The differences in planet radii between our reproduced models and published results in \citet{Zeng2016} are barely noticeable despite the difference in the surface layer setting.

\begin{figure}
  \centering
 \includegraphics[width=\columnwidth]{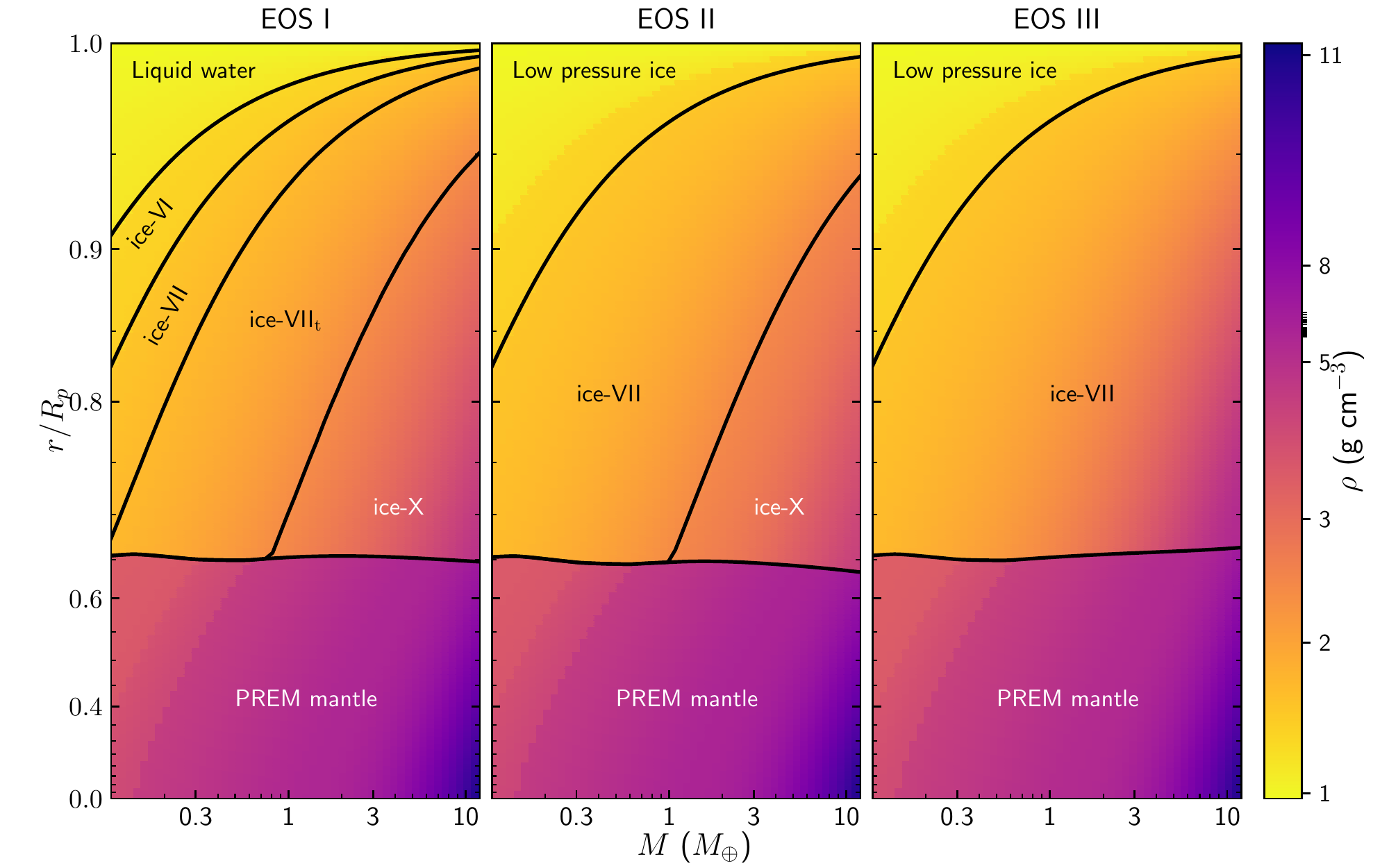}
\caption{
The planet interior density map for 50 wt\% water/rock planet model with a total mass between 0.1 and 10 $M_\oplus$.  The y-axis indicates the fraction of the total planet radius in an exponential scale, which is chosen to highlight the relatively thin water layer.  Black solid curves mark the phase boundary between different compositions or phases.  Because of the large density gap between the ice and rock, the colour scale is stretched near the two ends to better show the density gradient within each component.
}
\label{DM}
\end{figure}

To intuitively display the planet model setup, and to illustrate the application of the newly identified water ice phase, Fig.~\ref{DM} compares the internal structure one would find on a large, core-free, Ganymede-like planet for a range of masses between 0.1 and 10 times that of Earth.  An example of a system of planets with roughly these properties is TRAPPIST-1.  TRAPPIST-1f, in particular, is roughly one Earth mass and early measurements suggested water content as high as 50\%---though recent updates to the mass lower this value to 10-15\% \citep{Unterborn2018NA, Unterborn2018RNAAS, Agol2021}.

Three panels from left to right show planet models using the ice EOS~I, II, and III described in Fig.~\ref{EOS_Ice} respectively.  Black solid curves mark the phase boundary between different compositions or phases.  The ice in the 2.1--88\,GPa  pressure range studied by \citet{Grande:2019} can occupy a substantial planet volume for water-rich planets.  The low-pressure section ($P<2.2$\,GPa ) are composed of liquid water and ice-VI in all three panels.  Because it is not an exact duplicate of the setup in~\citet{Zeng2016}, who used ice-Ih, ice-III, ice-V, and ice-VI along the melting line, the phase boundary is not marked and the phase is simply named ``low pressure ice'' to avoid confusion.

The new ice-VII$_\text{t}$ phase identified in \citet{Grande:2019} would primarily appear in water-rich planets with masses between 0.1 and 10 Earth masses.  Ice-VII$_\text{t}$ first appears at the water/mantle boundary at roughly the mass of Mars.  At one Earth mass, ice-VII$_\text{t}$ dominates the water layer, comprising nearly 20\% of the planet radius, and a small layer of ice-X could exist at the water/mantle boundary.

\section{The Mass-Radius Relationship}

\begin{figure*}
\includegraphics[width=\textwidth]{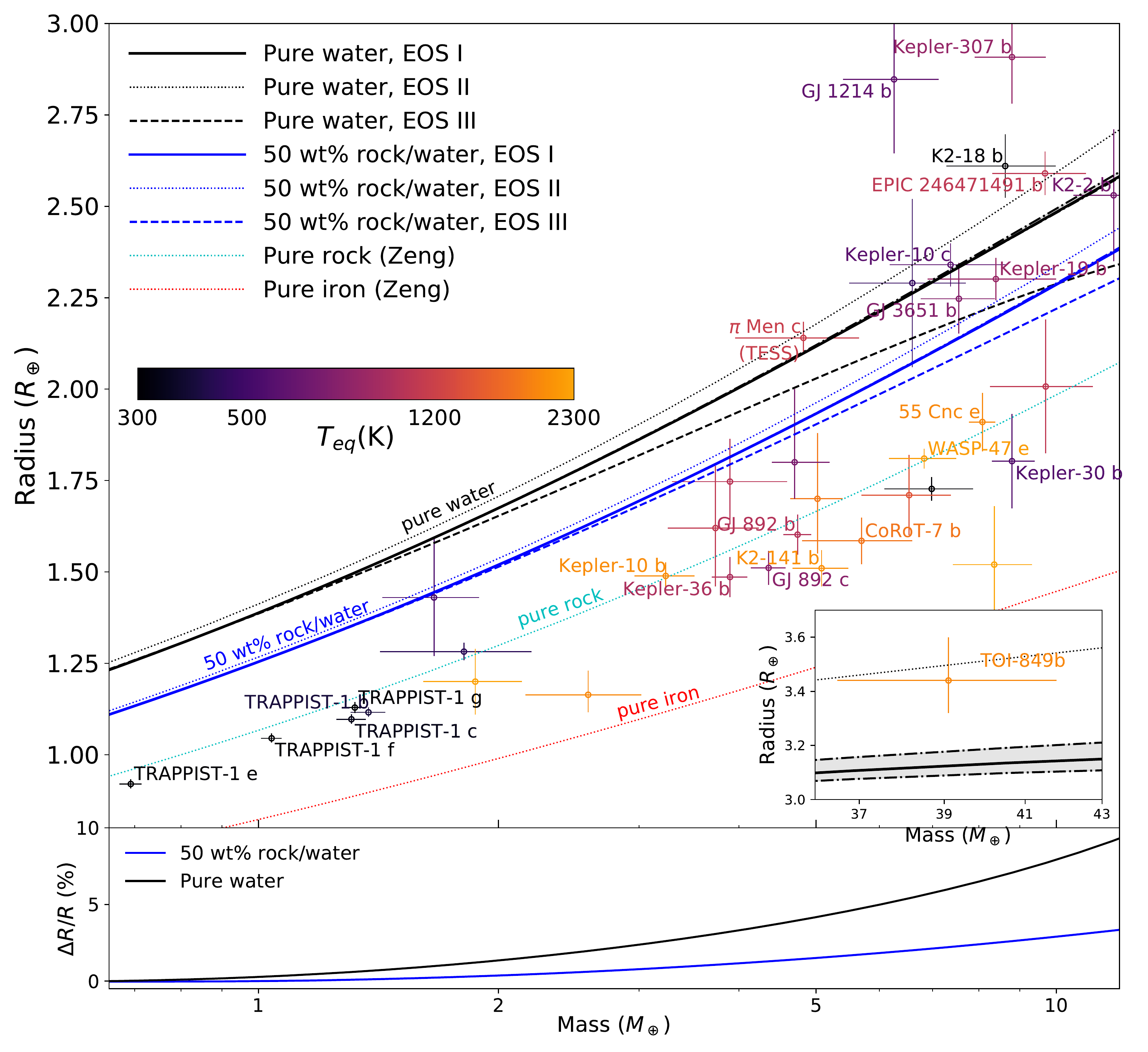}
\caption{\textbf{Upper:} Impact of ice EOS measurement on modelled planet mass-radius curves.
Results using the ice EOS measured by \citet{Grande:2019} are shown by the solid mass-radius curves.
Dotted curves show the relation using the model in \citet{Zeng2016}, whose temperature increases with pressure.
The dashed curves applied the EOS measured by \citet{Frank2004} at 300\,K.  Without the temperature dependence, it is a direct comparison with the solid line to show the impact of ice EOS measurements.
If it is possible to consider the temperature profile properly, planet model using recent ice EOS measurements would suggest a larger radius than the result shown in the dotted line.
Planets whose radii and masses are measured to better than $\sim{}10\%$ and $\sim{}20\%$ respectively are plotted and colour-coded by their surface temperatures.  \textbf{Inset:} Mass-radius relation near TOI-849b.
\textbf{Lower:} The percentage difference in radius between solid lines and dashed lines. 
}
\label{RM}
\end{figure*}

Fig.~\ref{RM} shows the mass-radius curve of pure water planet model and 50 wt\% water/rock planet using the method and EOSs described in the previous two sections.  The pure rock and pure iron planet models from~\citet{Zeng2016} are also shown for reference.  To estimate the impact of the uncertainty in the EOS parameters on the mass-radius curve, we repeat the calculation by randomly drawing 1000 sets of EOS fitting parameters from the MCMC posteriors of \cite{Grande:2019}.  The range of 1-$\sigma$ uncertainty is marked as a grey band that is bounded by dash-dotted lines.
Nevertheless, this statistical uncertainty is too small to see since the widths of the bands are thinner than the thicknesses of the solid lines.

Fig. \ref{RM} shows that planets modelled following the prescription of \citet{Zeng2016} are larger than those that use the 300\,K isotherm EOS from \citet{Frank2004} because of their estimate of temperature effects (planets using EOS~II are larger than planets using EOS~III).  
This difference is even noticeable at the low-mass end where extrapolating the EOS is not required.  Such a difference would be even more dramatic if the thickness of the atmosphere were not negligible.
Because most of planet's interior is hotter than the 300\,K that is assumed for the ice layer in this work, the mass-radius curve shown in solid line likely underestimates the true mass-radius curve that would arise from thermal expansion of high-pressure ice.
To gain insight into the effects of the new EOS compared to existing measurements, it is best to compare models that use EOS~I and those that use EOS~III because with these EOS measurements were done under the same physical conditions -- the only difference between them being the estimated model parameters.

The significant increase of bulk modulus of ice-X measured by \citet{Grande:2019} yields an extrapolated density of ice at high pressure that is smaller than suggested by \citet{Frank2004}.  The inset plot in Fig.~\ref{RM} shows the planet radius ratio of models using EOS~I to those using EOS~III in the same planet mass range.  The ice EOS of \citet{Grande:2019} implies larger planets for a given mass and the difference increases rapidly with the mass of the ice layer.  Assuming that the EOS~II used by \citet{Zeng2016} correctly estimates the thermal effect, then a similar planet model that uses the new EOS would produce larger planet radii than the result using EOS~II (shown by the dotted line).

In contrast to the simplified mantle-ice two-layer planet model shown in Fig.~\ref{RM}, an iron-rich core is expected to exist in the center of most terrestrial planets \citep[e.g.][]{Sotin2007}.  When keeping a planet's total mass, replacing a fraction of its mantle by an iron core would increase the planet's density and shrink the planet's size.  Thus, the pressure at the mantle-ice boundary is higher for planets with an iron core.  Since the difference between ice EOSs increases with pressure, the impact of applying different EOS parameters would lead to even larger differences in predicted planet radii.  As an example, we consider a 300K isothermal planet with 1.7 \MEarth\ iron core, 3.3 \MEarth\ rock mantle, and 5 \MEarth\ Ice.  This iron/mantle mass ratio is similar to that of the Earth.  The model using the ice EOS~I suggests a planet-size of 2.201 \REarth\ for such composition, which is 3.8\% larger than the 2.121 \REarth\ resulting from EOS~III.  By comparison, the planet radius only increases 3.0\% (2.280 \REarth\ vs 2.213 \REarth) for a 50\% water/50\% rock 10 \MEarth\ planet.

To quantify the change in planet radii, we compare simulated mass-radius curves with the masses and radii of observed exoplanets, as well as their measurement uncertainties.  We download all planets whose planet mass and radius limit tag are 0 from the NASA exoplanet archive and merge the list with TEPCat \citep{TEPCat}.  From the list, we pick planets with relative mass and radius uncertainty smaller than 50\%, and remove planets heavier than 30 $M_\oplus$ or larger than 5 $R_\oplus$.  The mass and radius of those planets are updated with recent data, e.g. TTV masses calculated by \citet{Hadden2017} and planet radius derived in \citet{Fulton2018} or \citet{Berger2018} are used when available.  After these selections, those planets that have mass and radius uncertainties less than 25\% are listed in the Table~\ref{tab:MR} with their reference, and are shown in Fig.~\ref{RM}.  Planet names are labelled only for those which have high measurement precision or are otherwise noteworthy.

The planets are colour-coded with their equilibrium temperature.  If the equilibrium temperature is not available in the literature, we estimate the $T_{eq}$ assuming albedo is $0$.  Therefore, if insolation $I_0$ is available, 
\begin{equation}
  \label{eq:Teq1}
  T_{eq} = ( I_0 / 4 \sigma )^{(1/4)},
\end{equation}
where $\sigma$ is Stefan--Boltzmann constant.  Or, 
\begin{equation}
  \label{eq:Teq2}
  T_{eq} = T_{\text{eff}}\sqrt{R_{\star}/(2a)},
\end{equation}
where $T_{\text{eff}}, R_{\star}$ is the effective temperature and radius of the host star and $a$ is the semimajor axis of the planet.

The mass-radius relationship in Fig.~\ref{RM} shows that the systematic differences between previous EOS measurements and our own are comparable to uncertainties in planet size measurements.  This figure also highlights the classes of planets that are most affected by the recent EOS measurements. 

The differences between the different EOSs are more drastic at higher masses such as in the case of TOI-849 b shown in the inset of Fig.~\ref{RM}.  With mass of $39.1 ^{+2.7}_{-2.6} M_\oplus$ and density of $5.2 ^{+0.7}_{-0.8}$ g cm$^{-3}$, TOI-849 b shows that planets could exist in the Neptune desert without large fractions of their mass in a gas envelope \citep{TOI-849}.  The difference in radii between a pure water planet with EOS~I and EOS~II at TOI-849's mass is 0.39 $R_\oplus$, which is larger than the 1-$\sigma$ observed uncertainty $\left(0.28 R_\oplus\right)$.  In addition, the 1-$\sigma$ uncertainty in planet size that results from the \citet{Grande:2019} measurement uncertainty (the grey band) is also comparable to the observational uncertainty.  These results highlight the need for improved EOS measurements when making first-order inferences for the interior structure of Neptune-like exoplanets.  However, we note that current EOSs are not likely suitable for extrapolation to these masses and temperatures, and the interior of gas giants are likely composed of more exotic mixtures of elements.

At lower mass, many planets -- especially those larger than $\sim$2 $R_\oplus$ -- likely have a sizable atmosphere that contributes to their radius \citep{Fulton2017}.  This additional layer adds another component to an already degenerate model.  Nevertheless, the new EOS measurements provide a more accurate, minimum contribution that an atmosphere can make to the planet radius for planets that lie above the 100\% water model.

\begin{figure}
\includegraphics[width=\columnwidth]{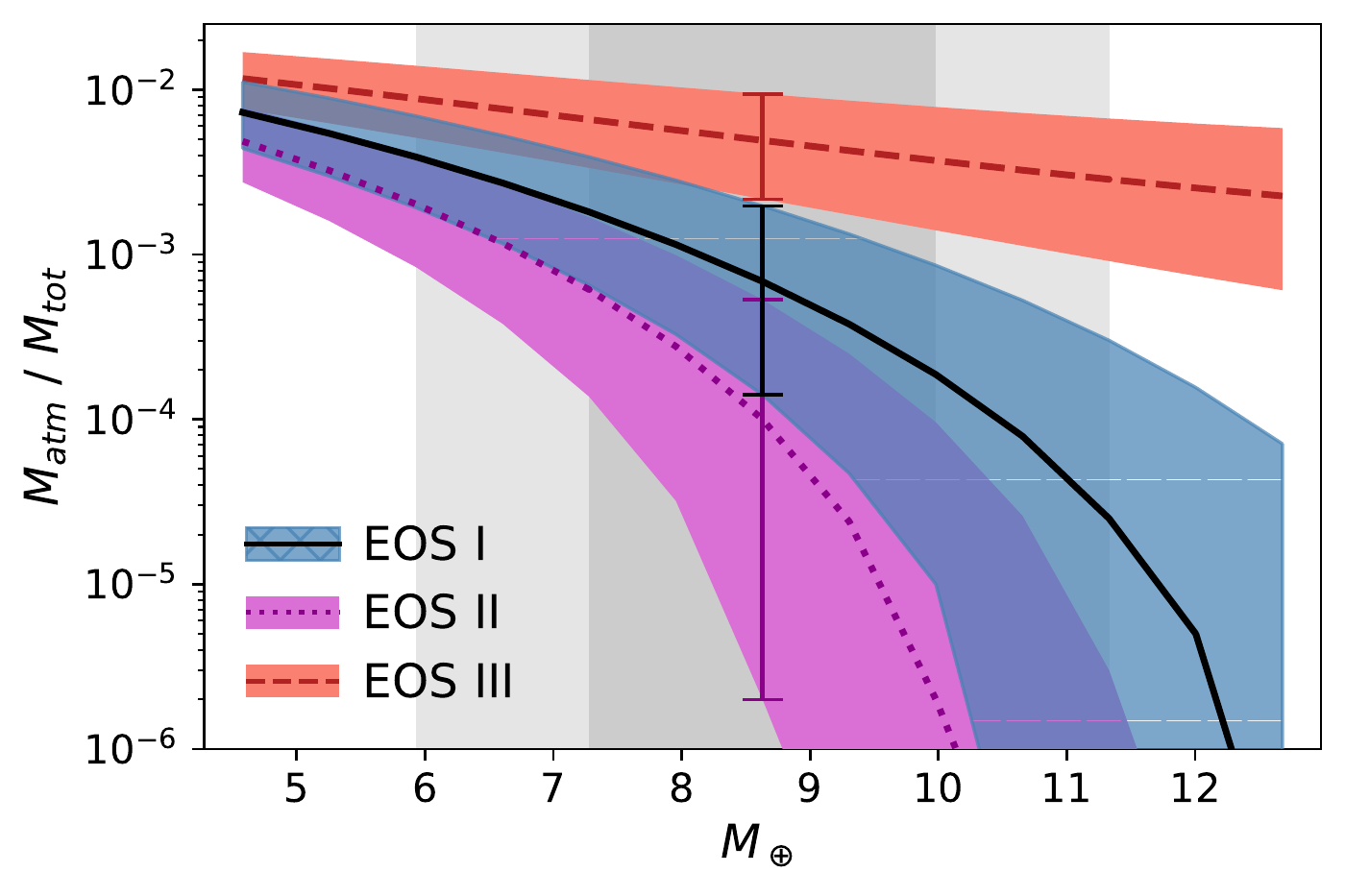}
\caption{Impact of ice EOS measurements on inferences of minimum atmospheric weight for K2-18b. The three EOS shown are labeled the same as Fig.~\ref{RM}---EOS~I: \citet{Grande:2019}, EOS~II: \citet{Zeng2016}, EOS~III: \citet{Frank2004}. Lines show the fraction of planet mass in an atmosphere needed to reproduce the radius of K2-18b across a range of likely masses.  Planets are modeled as pure water with a 3.0 g/mol ideal gas atmosphere and a equilibrium temperature of 255 K.  The grey areas show the one and two sigma mass bounds of K2-18b.  Planets with varying atmosphere are run at each half step in sigma.  The shaded regions show the atmosphere needed to reproduce the radius of K2-18b to within one sigma for each EOS.  One error bar is shown at 8.63 $M_\oplus$.  Planet parameters from \citet{K2-18b}.  
}
\label{k2}
\end{figure}

As an example, we consider an unlikely water and atmosphere model to demonstrate the impact of the EOS on inferences of minimum atmosphere mass.  K2-18 b, which is clearly above the pure water lines on Fig.~\ref{RM}, has a mass of 8.63 $\pm$ 1.35 $M_\oplus$ \citep{K2-18}, a radius of 2.61 $\pm$ 0.087 $R_\oplus$, and confirmed water vapor in its atmosphere \citet{K2-18b}.  The planet is in the habitable zone with an equilibrium temperature near 255 K.  \citet{Madhusudhan2020} exhaustively models K2-18 b.  However, here we opt to use a two layer model of pure water and an ideal gas to find the absolute minimum amount of atmosphere needed to match observations.  We use the model presented in Section \ref{sec:model} with an outer temperature of 255 K.  The ideal gas atmosphere has a molecular weight of 3.0 g/mol and an adiabatic temperature gradient.  We model planets with varying atmosphere fractions with total mass ranging from $\pm 3 \sigma$ of K2-18 b's observed mass at half integer steps.  In Fig.~\ref{k2}, we show the fraction of mass needed in the atmosphere to reproduce the observed radius of the planet assuming a fully opaque atmosphere.  The simulated radii of the planets with atmosphere mass fractions greater than $10^{-6}$ are all within 0.1\% of the planet's observed radius and its $\pm 1 \sigma$ uncertainties.

As seen on Fig.~\ref{k2}, for EOS~I from \citet{Grande:2019} K2-18 b would need a minimum H/He ideal gas mass fraction of $6.98\times 10^{-4}$ to reproduce it's observed mean radius within 0.002\%.  EOS~I requires 6.9 times more atmosphere by mass than EOS~II \citep{Zeng2016} in order for the planet to have a radius of 2.61 $R_\oplus$, and it requires 7.2 times less mass than EOS~III \citep{Frank2004} at a mass of 8.63 $M_\oplus$ for the same result.  The difference in inferred atmosphere fraction between the EOSs increases with observed mass.

These calculations, using the new EOS measurements for water show that ongoing improvements in exoplanet mass and radius measurements require similar improvements in the EOS measurements of planetary materials.  Our ability to model the interior planet structure and dynamics will be increasingly limited by our knowledge of the properties of the planetary material under high-pressure and high-temperature conditions.

\section{Conclusions}

We show that the larger bulk modulus measured by \citet{Grande:2019} suggests smaller ice-X density compared to the result of \citet{Frank2004} under the same physical conditions.  Thus, the size of water rich planets are larger than previous models suggest (such as \citet{Zeng2016}).  At 10 Earth masses, ice-X is the prevailing phase comprising 20\% of the radius, while most of the remaining 10\% is the new phase, ice-VII$_\text{t}$.  The larger simulated planet radius for a given mass either requires smaller atmospheric contribution to the volume of many exoplanets or limits their water content. Instead of suggesting a more accurate mass-radius relation for water-rich planets, our results reveal the large uncertainty in the mass-radius relation caused by disagreement between experimental measurements of the properties of planet-forming material.  These variations will propagate into uncertainties in planet structure and bulk composition.  These issues should receive more attention and discussion.

At this time, the observational uncertainty of planet mass and radius measurements is comparable to the variation of the modelled planet radii that comes from a lack of consensus in measured EOSs.  The differences between EOS measurements produce planets whose differences in radii are larger than what one would predict from the stated measurement uncertainties.  For example, assuming the stated experimental uncertainties in \citet{Grande:2019} are accurate, a 5 M$_\oplus$ pure water planet model adopting EOS~I would produce a 50-$\sigma$ disagreement in the predicted planet radii compared with the model adopting EOS~III.  For a 50/50 water/rock composition at the same planet mass, the disagreement is 63-$\sigma$. 
As the observational precision of exoplanet properties improves, this issue will be exacerbated, and our ability to constrain planet compositions will be increasingly limited by the laboratory work.  The careful study of the properties of ice, including the phase diagram, the equations of state, Gr\"{u}neisen parameter, and the adiabatic bulk modulus, under the anticipated temperature and pressure conditions, is crucial for constructing reliable planet models and inferring planet composition.  

This work at room temperature has given insights into the importance of reevaluating planetary models as new studies are performed on relevant materials.  Future work on high pressure ice at elevated temperatures are needed to continue improving the accuracy of these models which, at best, are limited to using primarily theoretical results to calculate thermal effects within planets (e.g. \citet{Sotin2007,Grimm2018,Haldemann2020}).  Future work on extending the high pressure melt-line, determining the thermal expansion at higher pressures, and measuring the Clayperon slope of the Ice-VIIt to Ice-X transition are all needed to improve the accuracy of planet models.  Recent measurements appear for ices II, III, V, and VI in \citet{Journaux2020}, which are used to model planets in \citet{Journaux2020a}.  Future measurements will not only impact interior structure models, but also our understanding of heat transfer and geodynamics in water-rich planets.

\section*{Acknowledgements}

C.H. acknowledge support from NASA grants NNX16AK32G and NNX16AK08G.  Portions of this work were performed at HPCAT (Sector 16), Advanced Photon Source (APS), Argonne National Laboratory. HPCAT operations are supported by DOE-NNSA’s Office of Experimental Sciences.  The Advanced Photon Source is a U.S. Department of Energy (DOE) Office of Science User Facility operated for the DOE Office of Science by Argonne National Laboratory under Contract No. DE-AC02-06CH11357. C.H. thanks the discussion with Shichun Huang and Keith Lawler.  We also thank the referee for providing constructive comments and suggestions.




\bibliographystyle{mnras}
\bibliography{references} 

\newpage
\onecolumn
\begin{longtable}{cccc}
  \caption{List of potential terrestrial planets with mass and radius measurement accuracy better than 25\%} 
  \label{tab:MR}\\
  \hline
  Planet & $T_{eq}$ (K) & $M_p/M_\oplus$ & $R_p/R_\oplus$ \\
  \hline
  \hline
  \endfirsthead

  {\tablename\ \thetable\ -- \textit{Continued.}}\\
  \hline
  Planet & $T_{eq}$ (K) & $M_p/M_\oplus$ & $R_p/R_\oplus$ \\
  \hline
  \hline
  \endhead
  \hline
  \endfoot
  55 Cnc e~\citep{55Cnc} & 1958 & $8.59^{+0.43}_{-0.43}$ & $1.95^{+0.04}_{-0.04}$\\ 
  $\pi$ Men c~\citep{piMenc} & 1170 & $4.82^{+0.84}_{-0.86}$ & $2.14^{+0.04}_{-0.04}$ \\
  CoRoT-7 b~\citep{CoRoT7} & 1756 & $5.7^{+0.9}_{-0.9}$ & $1.59^{+0.06}_{-0.06}$ \\
  EPIC 246471491 b~\citep{EPIC246471491} & 1089 & $9.68^{+1.21}_{-1.37}$ & $2.59^{+0.06}_{-0.06}$ \\
  EPIC 246471491 c~\citep{EPIC246471491} & 741 & $15.68^{+2.28}_{-2.13}$ & $3.53^{+0.08}_{-0.08}$ \\
  GJ 436 b~\citep{GJ436} & 670 & $22.1^{+2.3}_{-2.3}$ & $4.17^{+0.17}_{-0.17}$\\
  GJ 892 b~\citep{GJ892} & 1015 & $4.74^{+0.19}_{-0.19}$ & $1.60^{+0.06}_{-0.06}$\\
  GJ 892 c~\citep{GJ892} & 782 & $4.36^{+0.22}_{-0.22}$ & $1.51^{+0.05}_{-0.05}$\\
  GJ 1132 b~\citep{GJ1132b}~\citep{GJ1132bR} & 529 & $1.66^{+0.23}_{-0.23}$ & $1.43^{+0.16}_{-0.16}$\\
  GJ 1214 b~\citep{GJ1214} & 604 & $6.26^{+0.86}_{-0.86}$ & $2.85^{+0.20}_{-0.20}$ \\
  GJ 3053 b~\citep{LHS1140} & 235 & $6.98^{+0.89}_{-0.89}$ & $1.73^{+0.03}_{-0.03}$ \\
  GJ 3053 c~\citep{LHS1140} & 438 & $1.81^{+0.39}_{-0.39}$ & $1.28^{+0.02}_{-0.02}$ \\
  GJ 3470 b~\citep{GJ3470} & 593 & $13.9^{+1.5}_{-1.5}$ & $4.57^{+0.18}_{-0.18}$ \\
  GJ 3651 b~\citep{GJ3651} & 760 & $7.55^{+0.83}_{-0.79}$ & $2.25^{+0.10}_{-0.10}$ \\
  K2-2 b~\citep{K2-2} & 690 & $11.8^{+1.3}_{-1.3}$ & $2.53^{+0.18}_{-0.18}$ \\
  K2-3 b~\citep{K2-3} & 463 & $6.6^{+1.1}_{-1.1}$ & $2.29^{+0.23}_{-0.23}$ \\
  K2-18 b~\citep{K2-18b} & 265 & $8.63^{+1.4}_{-1.4}$ & $2.61^{+0.09}_{-0.09}$ \\
  K2-38 b~\citep{K2-38} & 1184 & $12^{+2.9}_{-2.9}$ & $1.55^{+0.16}_{-0.16}$ \\
  K2-66 b~\citep{K2-66} & 1372 & $21.3^{+3.6}_{-3.6}$ & $2.49^{+0.34}_{-0.24}$ \\
  K2-95 b~\citep{K2-95} & 420 & $11.0^{+2.7}_{-2.7}$ & $3.47^{+0.78}_{-0.53}$ \\
  K2-96 b~\citep{K2-96} & 1800 & $5.02^{+0.38}_{-0.38}$ & $1.70^{+0.18}_{-0.15}$ \\
  K2-96 c~\citep{K2-96} & 500 & $9.8^{+1.3}_{-1.24}$ & $3.01^{+0.42}_{-0.28}$ \\
  K2-106 b~\citep{K2-106} & 2333 & $8.36^{+0.96}_{-0.94}$ & $1.52^{+0.16}_{-0.16}$ \\
  K2-110 b~\citep{K2-110} & 640 & $16.7^{+3.2}_{-3.2}$ & $2.60^{+0.10}_{-0.10}$ \\
  K2-131 b~\citep{K2-131} & 2776 & $6.5^{+1.6}_{-1.6}$ & $1.81^{+0.16}_{-0.12}$ \\
  K2-135 b~\citep{K2-135} & 1114 & $3.74^{+0.5}_{-0.48}$ & $1.62^{+0.17}_{-0.16}$ \\
  K2-141 b~\citep{K2-141} & 2039 & $5.08^{+0.41}_{-0.41}$ & $1.51^{+0.05}_{-0.05}$ \\
  K2-155 b~\citep{K2-155} & 708 & $4.7^{+0.5}_{-0.3}$ & $1.8^{+0.2}_{-0.1}$ \\
  K2-155 c~\citep{K2-155} & 583 & $6.5^{+1.5}_{-0.5}$ & $2.6^{+0.7}_{-0.2}$ \\
  K2-229 b~\citep{K2-229} & 1960 & $2.59^{+0.43}_{-0.43}$ & $1.16^{+0.07}_{-0.05}$ \\
  K2-263 b~\citep{K2-263} & 470 & $14.8^{+3.1}_{-3.1}$ & $2.41^{+0.12}_{-0.12}$ \\
  K2-265 b~\citep{K2-265} & 1400 & $6.54^{+0.84}_{-0.84}$ & $1.71^{+0.11}_{-0.11}$ \\
  Kepler-4 b~\citep{K4} & 1650 & $24.5^{+3.8}_{-3.8}$ & $4.08^{+0.11}_{-0.11}$ \\
  Kepler-10 b~\citep{K10, Fulton2018}& 2000 & $3.24^{+0.28}_{-0.28}$ & $1.49^{+0.04}_{-0.04}$ \\
  Kepler-10 c~\citep{K10, Fulton2018} & 580 & $7.37^{+1.32}_{-1.19}$ & $2.34^{+0.06}_{-0.06}$ \\
  Kepler-11 d~\citep{Hadden2017, Fulton2018} & 730 & $6.8^{+0.7}_{-0.8}$ & $3.38^{+0.10}_{-0.10}$ \\
  Kepler-11 e~\citep{Hadden2017, Fulton2018} & 650 & $6.7^{+1.2}_{-1.0}$ & $4.04^{+0.11}_{-0.11}$ \\
  Kepler-19 b ~\citep{Kepler19, Fulton2018} & 860 & $8.4^{+1.6}_{-1.5}$ & $2.30^{+0.06}_{-0.06}$ \\
  Kepler-20 b ~\citep{K20, Fulton2018} & 1105 & $9.7^{+1.4}_{-1.4}$ & $2.01^{+0.18}_{-0.18}$ \\
  Kepler-20 c ~\citep{K20, Fulton2018} & 772 & $12.8^{+2.2}_{-2.2}$ & $2.88^{+0.13}_{-0.13}$ \\
  Kepler-26 b ~\citep{Hadden2017, Berger2018} & 427 & $5.1^{+0.6}_{-0.7}$ & $3.19^{+0.10}_{-0.09}$ \\
  Kepler-26 c ~\citep{Hadden2017, Berger2018} & 381 & $6.1^{+0.7}_{-0.7}$ & $2.98^{+0.28}_{-0.23}$ \\
  Kepler-30 b ~\citep{Hadden2017, Fulton2018} & 558 & $8.8^{+0.6}_{-0.5}$ & $1.80^{+0.13}_{-0.13}$ \\
  Kepler-33 e ~\citep{Hadden2017, Fulton2018} & 794 & $5.5^{+1.2}_{-1.1}$ & $3.48^{+0.11}_{-0.11}$ \\
  Kepler-33 f ~\citep{Hadden2017, Fulton2018} & 729 & $9.6^{+1.7}_{-1.8}$ & $3.94^{+0.13}_{-0.13}$ \\
  Kepler-36 b ~\citep{Hadden2017, Fulton2018} & 978 & $3.9^{+0.2}_{-0.2}$ & $1.49^{+0.06}_{-0.06}$ \\
  Kepler-36 c ~\citep{Hadden2017, Fulton2018} & 928 & $7.5^{+0.3}_{-0.3}$ & $3.96^{+0.14}_{-0.14}$ \\
  Kepler-48 c ~\citep{Marcy2014, Fulton2018} & 784 & $14.6^{+2.3}_{-2.3}$ & $2.56^{+0.07}_{-0.07}$ \\
  Kepler-49 b ~\citep{Hadden2017, Berger2018} & 598 & $8^{+1.9}_{-1.6}$ & $2.61^{+0.08}_{-0.08}$ \\
  Kepler-60 b ~\citep{Hadden2017, Fulton2018} & 1247 & $3.7^{+0.6}_{-0.6}$ & $1.95^{+0.24}_{-0.24}$ \\
  Kepler-60 c ~\citep{Hadden2017, Fulton2018} & 1158 & $2.0^{+0.3}_{-0.5}$ & $2.09^{+0.22}_{-0.22}$ \\
  Kepler-60 d ~\citep{Hadden2017, Fulton2018} & 1052 & $3.9^{+0.7}_{-0.6}$ & $1.75^{+0.12}_{-0.12}$ \\
  Kepler-78 b ~\citep{K78} & 2250 & $1.87^{+0.27}_{-0.26}$ & $1.20^{+0.09}_{-0.09}$ \\
  Kepler-93 b ~\citep{K93, Fulton2018}& 1037 & $4.0^{+0.7}_{-0.7}$ & $1.63^{+0.06}_{-0.06}$ \\
  Kepler-94 b ~\citep{Marcy2014, Fulton2018} & 1068 & $10.8^{+1.4}_{-1.4}$ & $3.05^{+0.12}_{-0.12}$ \\
  Kepler-95 b ~\citep{Marcy2014, Fulton2018} & 1009 & $13.0^{+2.9}_{-2.9}$ & $3.12^{+0.09}_{-0.09}$ \\
  Kepler-99 b ~\citep{Marcy2014, Fulton2018} & 890 & $6.2^{+1.3}_{-1.3}$ & $1.81^{+0.14}_{-0.14}$ \\
  Kepler-102 e ~\citep{Marcy2014, Fulton2018} & 590 & $8.9^{+2.0}_{-2.0}$ & $2.41^{+0.14}_{-0.14}$\\
  Kepler-131 b ~\citep{Marcy2014, Fulton2018} & 778 & $16.1^{+3.5}_{-3.5}$ & $2.06^{+0.05}_{-0.05}$ \\
  Kepler-177 b ~\citep{Hadden2017, Fulton2018} & 691 & $5.4^{+1.0}_{-0.9}$ & $4.50^{+0.39}_{-0.39}$ \\
  Kepler-238 f ~\citep{K238, Fulton2018}& 653 & $13.5^{+2.9}_{-2.5}$ & $3.48^{+0.41}_{-0.41}$ \\
  Kepler-289 d ~\citep{K289}& 2500 & $4.0^{+0.9}_{-0.9}$ & $2.68^{+0.17}_{-0.17}$ \\
  Kepler-307 b ~\citep{Hadden2017, Fulton2018} & 838 & $8.8^{+0.9}_{-0.9}$ & $2.91^{+0.13}_{-0.13}$ \\
  Kepler-307 c ~\citep{Hadden2017, Fulton2018} & 777 & $3.9^{+0.7}_{-0.7}$ & $2.72^{+0.17}_{-0.17}$ \\
  Kepler-406 b ~\citep{Marcy2014, Fulton2018} & 1482 & $6.4^{+1.4}_{-1.4}$ & $1.45^{+0.04}_{-0.04}$ \\
  Kepler-454 b ~\citep{K454}& 912 & $6.8^{+1.4}_{-1.4}$ & $1.84^{+0.06}_{-0.06}$ \\
  NGTS-4 b ~\citep{NGTS4}& 1650 & $20.8^{+3.4}_{-3.4}$ & $3.18^{+0.27}_{-0.27}$ \\
  TOI-849 b ~\citep{TOI-849}& 1932 & $39.1^{+2.7}_{-2.6}$ & $3.44^{+0.16}_{-0.12}$ \\
  TRAPPIST-1 b ~\citep{newTRAPPIST1} & 392 & $1.374^{+0.069}_{-0.069}$ & $1.116^{+0.014}_{-0.012}$ \\
  TRAPPIST-1 c ~\citep{newTRAPPIST1} & 335 & $1.308^{+0.056}_{-0.056}$ & $1.097^{+0.014}_{-0.012}$ \\
  TRAPPIST-1 d ~\citep{newTRAPPIST1} & 282 & $0.388^{+0.012}_{-0.012}$ & $0.788^{+0.011}_{-0.010}$ \\
  TRAPPIST-1 e ~\citep{newTRAPPIST1} & 246 & $0.692^{+0.022}_{-0.022}$ & $0.920^{+0.013}_{-0.012}$ \\
  TRAPPIST-1 f ~\citep{newTRAPPIST1} & 215 & $1.039^{+0.031}_{-0.031}$ & $1.045^{+0.013}_{-0.012}$ \\
  TRAPPIST-1 g ~\citep{newTRAPPIST1} & 195 & $1.321^{+0.038}_{-0.038}$ & $1.129^{+0.015}_{-0.013}$ \\
  TRAPPIST-1 h ~\citep{newTRAPPIST1} & 169 & $0.326^{+0.020}_{-0.020}$ & $0.755^{+0.014}_{-0.014}$ \\
  WASP-47 d ~\citep{WASP47} & 960 & $13.1^{+1.5}_{-1.5}$ & $3.58^{+0.05}_{-0.05}$ \\
  WASP-47 e ~\citep{WASP47} & 2200 & $6.83^{+0.66}_{-0.66}$ & $1.81^{+0.03}_{-0.03}$ \\

\end{longtable}
\clearpage
\twocolumn

\bsp	
\label{lastpage}
\end{document}